\documentstyle[aps,preprint]{revtex}
\begin{document}
\setcounter{page}{0}
\def\footnoterule{\kern-3pt \hrule width\hsize \kern3pt}
\title{SYMPLECTIC REDUCTION AND SYMMETRY ALGEBRA IN BOUNDARY CHERN-SIMONS  
THEORY\thanks
{This work is supported in part by funds provided by the U.S.
Department of Energy (D.O.E.) under cooperative 
research agreement \#DF-FC02-94ER40818.}}

\author{ Phillial  Oh\footnote{Permanent address: Department of Physics,
Sung Kyun Kwan University, Suwon 440-746, Korea;
Email address: ploh@ctpa03.mit.edu
} and Mu-In Park
\footnote{Email address: {mipark@physics.sogang.ac.kr}}
}
\address{Center for Theoretical Physics \\
and Laboratory for Nuclear Science \\
Massachusetts Institute of Technology \\
Cambridge, Massachusetts 02139 \\
{~}}


\maketitle

\thispagestyle{empty}

\begin{abstract}
We derive the Kac-Moody algebra and Virasoro algebra in Chern-Simons theory 
with boundary by using the symplectic reduction method and the Noether
procedures.
\end{abstract}

\vspace*{\fill}

\section{Introduction}

There has been  vast interest in Chern-Simons 
theory \cite{ref:Des} with boundary \cite{ref:Bal} in 
diverse areas of physics,
especially in Chern-Simons gravity theory \cite{ref:Ach}
because of a  relation with black hole entropy \cite{ref:Car,ref:Srt}. 
Recently, Ba\~nados {et. al} \cite{ref:Ban1}
further argued that the Kac-Moody and  Virasoro algebras 
on the boundary of the black hole play a crucial role
in understanding the statistical origin of the 
BTZ black hole entropy \cite{ref:BTZ}. However, the derivation of these
algebras was based on some assumptions about boundary charges.
Also the relation with standard Noether procedures and
constraint analysis  were not clear.

In this Letter,  we rederive these algebras by applying 
the symplectic method \cite{ref:Fad} and the Noether procedure.
We shall identify the correct symplectic structure on the boundary,
calculate the Poisson bracket between gauge and diffeomorphism
charges, and  obtain the central terms in the
Kac-Moody and Virasoro algebras.

\section{Symplectic Structure}

Let us start from the Chern-Simons Lagrangian on the disc ${D}$
\begin{equation}
L=\frac{\kappa}{2 \pi} \int _{D} d^2 x \epsilon^{\mu \nu \rho} \left< A_{\mu} 
\partial_{\nu} A_{\rho} +\frac{2}{3} A_{\mu} A_{\nu} A_{\rho}\right>,
\label{eq:1}
\end{equation}
where $\left<\cdots\right>$ denotes trace.
Up to a boundary term, (\ref{eq:1}) 
can be put into the canonical form with the
Lagrangian
\begin{equation}
L=\frac{\kappa}{4 \pi}\int_{D} d^2 x \epsilon^{ij} ( A_i^a \dot{A}^{a}_{j}
-A_0^a F_{ij}^a).
\label{eq:2}
\end{equation}
(Here, $\epsilon^{012} \equiv \epsilon^{12}\equiv 
1,~ A_i =A^a_i t_a,~ F_{ij}=F_{ij}^a t_a,~F_{ij}^a=\partial_i A^a_j 
-\partial_j A^a _i +f^{abc} A_i^b A_j^c$, and the group generators $t^a$ 
satisfy $[t^a, t^b]=f^{abc} t^c,~\left<t^a t^b\right>=-\frac{1}{2}
\delta^{ab}$.)
We shall take (\ref{eq:2}) as our starting point. 
Variation with respect to 
$A_0^a$ gives the Gauss' law constraint
\begin{equation}
F_{ij}^a =0.
\label{eq:3}
\end{equation}
We adopt symplectic method and first  solve the constraint explicitly 
\cite{ref:Fad}. The solution 
is the pure gauge 
\begin{equation}
A_i = g^{-1} \partial_i g.
\label{eq:4}
\end{equation}
Substitution into the Lagrangian (2) gives
\begin{eqnarray}
L&=&-\frac{\kappa}{2 \pi} \int_{D} d^2 x \epsilon^{ij} \left< g^{-1} 
\partial_i g \partial_t (g^{-1} \partial_j g ) \right> \nonumber \\
 &=&\frac{\kappa}{2 \pi} \int_{D} d^2 x \epsilon^{ij} \left< \partial_i 
g^{-1} \partial_j g g^{-1} \dot{g} \right> + \frac{\kappa}{2 \pi} 
\oint_{\partial
 {D}} d \varphi \left< g^{-1} \partial_{\varphi} g g^{-1} \dot{g} \right> 
\nonumber \\
&\equiv& L_B +L_S, 
\label{lagrangian}
\end{eqnarray}
where $\varphi$ denotes the angular coordinate on the boundary $\partial {D}$
of disc $D$, and $L_B$ and $L_S$ are the bulk and surface Lagrangians, 
respectively. Upon the parameterization of $g$ locally, 
it can be shown that the $L_B$ is also a surface term \cite{ref:Jac1}. 

Let us first compute the symplectic structure \cite{witten}. 
Lagrangian (\ref{lagrangian}) suggests the following 
canonical 1-form
\begin{equation}
\Theta = \frac{\kappa}{2 \pi} \int_{D} d^2 x 
~\epsilon^{ij}\left< \partial_i g^{-1} 
\partial_j g g^{-1} d g \right> + \frac{\kappa}{2 \pi} \oint_{\partial {D}} 
d \varphi \left< g^{-1} \partial_{\varphi} g g^{-1} d g \right>, 
\label{symp1}
\end{equation}
where $d g$ is the functional exterior derivative of $g$.
Then, a straightforward computation gives 
\begin{equation}
\Omega=d \Theta =\frac{\kappa}{2 \pi} \oint_{\partial {D}} d \varphi 
\left< d (g^{-1} \partial_{\varphi} g)\wedge g^{-1} d g \right>.
\label{eq:6}
\end{equation} 
The above symplectic structure (\ref{eq:6}) 
yields the following Poisson bracket 
\begin{equation}
\{ (g^{-1} \partial_{\varphi} g)_{AB}(\varphi),~ g_{CD} (\varphi') \}=
 \frac{\pi}{\kappa}\delta_{AD} g_{CB}\delta(\varphi-\varphi^\prime),
\label{eq:7}
\end{equation}
where indices $A,~B, \cdots$ denote the components of the matrix.
To be more explicit, we parameterize the group element by some local 
coordinates $\theta^a~(a=1, \cdots, \mbox{dim}~ G )$: 
$g\equiv 
g(\theta^a)$. Let us 
define \cite{ref:Bak}
\begin{eqnarray}
 g^{-1}(\varphi) \frac{\partial g(\varphi)}{\partial \theta^a(\varphi ')}= 
{C_a} ^b (\varphi) t_b \delta (\varphi -\varphi ').
\label{done}
\end{eqnarray}
Then 
\begin{equation}
\Omega=\frac{1}{2} \oint_{\partial D} d \varphi \oint_{\partial D} d 
\varphi' \omega_{ab}(\varphi, \varphi^\prime)
 d \theta^a (\varphi)\wedge  d \theta^b (\varphi ') 
\end{equation}
 with
\begin{eqnarray}
\omega_{ab}(\varphi, \varphi') =-\frac{\kappa}{4 \pi}
 \left[ \frac{\partial A_{\varphi}^c (\varphi)}{ \partial \theta ^a 
(\varphi^\prime)} {C_b}^c(\varphi) -
\frac{\partial A_{\varphi}^c(\varphi)}{\partial 
\theta^b (\varphi^\prime)} {C_a}^c (\varphi) \right],
\label{eq:90}
\end{eqnarray}
where $g^{-1} \partial_{\varphi} g = A_{\varphi}^a (\theta) t_a$.
In general, this 2-form is degenerate,
and does not possess an inverse. Further reduction must occur from $G$ to 
one of its coadjoint orbit $G/H$ \cite{ref:Bak}. For our purpose, 
we  just assume the symplectic reduction 
has been performed and denote the coordinates on $G/H$ by  
$\bar{\theta}^{\alpha}~
(\alpha=1, \cdots, \mbox{dim}~ G/H =\mbox{even})$. Then the 2-form 
(\ref{eq:90}) descends onto symplectic 2-form 
$\omega_{\alpha \beta} (\varphi, \varphi')$ on $G/H$, and the
Poisson bracket is defined by 
\begin{eqnarray}
\{ \bar{\theta}^{\alpha}(\varphi), \bar{\theta}^{\beta}(\varphi ') \}=
\omega^{\beta \alpha} (\varphi ', \varphi), 
\end{eqnarray}
where $\omega^{\alpha\beta}$ is the inverse of
$\omega_{\alpha\beta}$.
Fortunately,  we do not need 
an explicit expression in terms of
local coordinates neither for
 $\omega^{\alpha\beta}$  
nor for  $ {C_\alpha} ^b$ of (\ref{done}).
Starting from the reduced expression of (\ref{eq:6}) on $G/H$, we find
\begin{equation}
\{ A_{\varphi}^a (\bar{\theta}(\varphi)), g(\bar{\theta}(\varphi')) \}
=-\frac{2 \pi}{\kappa} \delta(\varphi -\varphi ') g(\varphi ') t^a. 
\end{equation}
With this we calculate the Poisson bracket for $A_{\varphi}^a$:
\begin{eqnarray}
\{A_{\varphi}^a(\varphi), A_{\varphi}^b (\varphi ') \}
& =&\frac{2 \pi} {\kappa}
 f^{acb} A_{\varphi}^c 
(\varphi) \delta(\varphi -\varphi ') +\frac{ 2 \pi} {\kappa} 
\partial_{\varphi} \delta (\varphi -\varphi ') \delta^{ab}
\nonumber\\
&=&\frac{2 \pi} {\kappa}\left
(D_\varphi\delta(\varphi -\varphi^\prime)\right)^{ab},
\label{poisson}
\end{eqnarray}
which is the Kac-Moody algebra in density form. 
($D_{\varphi}$ is the $\varphi$-th component of the 
covariant derivative $D^{ab}_i=\delta^{ab} \partial_i +f^{acb} A^c_i$.)

\section{Kac-Moody Algebra of Gauge Transformation}

We now consider the gauge transformation generated by 
$\delta g = g \lambda$. From (\ref{lagrangian}), we obtain
\begin{eqnarray}
\delta L &=& \frac{d}{dt}  \left[\frac{\kappa}{2 \pi} \int_{D} d^2 x 
\epsilon^{ij} \left< g^{-1} \partial_{j} g g^{-1} \partial_{i} g \lambda 
\right> -
\frac{\kappa}{2 \pi} \oint_{\partial{D}} d \varphi \left< g^{-1} \partial_
{\varphi} g \lambda
 \right> 
\right]  \nonumber \\
& \equiv & \frac{d X}{dt}.  
\end{eqnarray}
Then the Noether charge associated with this gauge transformation is given by 
\begin{eqnarray}
Q(\lambda)&=&\left< \frac{\partial L} {\partial \dot{g}} \delta g \right> -X 
\nonumber \\
& =&\frac{\kappa}{\pi} \oint_{\partial {D}} d \varphi \left< g^{-1} 
\partial_{\varphi} g 
\lambda \right> = -\frac{\kappa}{2 \pi} \oint_{\partial{D}} d 
\varphi A_{\varphi}^a \lambda^a.
\end{eqnarray}
Using (\ref{poisson}), we find that 
$Q(\lambda)$ satisfies the Kac-Moody algebra:
\begin{equation}
\{Q(\lambda), Q (\eta) \} =Q([\lambda, \eta])-\frac{\kappa}{\pi} 
\oint_{\partial {D}} d \varphi \left< \lambda \partial_{\varphi} \eta \right>,
\label{kacmoody}
\end{equation}
where $[\lambda, \eta]^a =f^{abc}\lambda^b \eta^c$.

\section{Virasoro algebra of Diffeomorphism}

In the derivation of the Kac-Moody algebra (\ref{kacmoody}), 
the existence of the
central term does not depend on what boundary condition one chooses
for $\lambda$'s. The $\lambda$'s  just have to be 
non-constant and  single-valued functions on the boundary. However,
in the derivation of the Virasoro algebra, the existence of central term
depends crucially on the boundary condition one imposes.
In the computation of the Virasoro charge, constraints can be imposed from
the beginning as in the Kac-Moody case, or after the Noether procedure
has been applied. We examine both cases here, because they give different
boundary conditions in general, even though the final expressions of the
charges are the same.
Let us start with the Lagrangian (2)  and study the response of $L$ to a 
 spatial and time-independent diffeomorphism ({\it Diff}):
\begin{eqnarray}
\delta_f x^{\mu} &=&-\delta^{\mu}_{~ i} f^i, \nonumber \\
\delta_f A^a _i &=&f^j \partial_j A^a_i +
(\partial_i f^j) A_j^a, \nonumber \\
\delta_f A^a _0 &=&f^j \partial_j A^a_0.
\label{vari}
\end{eqnarray}
Under (\ref{vari}), we find
\begin{eqnarray}
\delta_f L &=& \frac{\kappa}{4 \pi} \int_{D} d^2 x  \epsilon^{ij} 
\partial _k [ f^k A^a_i \dot{A}^a_j-f^k A^a_0 F^a_{ij}] \nonumber \\
&=&\frac{\kappa}{4 \pi} \oint_{\partial D} 
d \varphi f^r (A_r^a \dot{A}^a_{\varphi}
 -\dot{A}^a_r A^a_{\varphi}- A^a_0 \epsilon^{ij} F_{ij}^a).
\label{boundary}
\end{eqnarray} 
Now, we have two possible boundary conditions 
in order that there be {\it Diff} 
invariance, i.e., $\delta_f L=\frac{d}{dt} X$:
(a). $f^r |_{\partial D}=0$, (b). 
$A^a_r |_{\partial D}$ =constant, $\partial_r
A^a_{\varphi}|_{\partial D} =0$, and 
${A^a_0}|_{\partial D}\propto {A^a_\varphi
}|_{\partial D}$.

(a). This is the simpler boundary condition for the {\it Diff} 
invariance with $X=0$ 
and results in {\it Diff}  only {\it along} the circle ($\partial D$). 
The Noether 
charge for this {\it Diff} becomes
\begin{eqnarray}
Q(f)&=&\frac{\partial L}{\partial \dot{A}^a_i} \delta_f A^a_i 
\nonumber \\
&=&\frac{\kappa}{4 \pi} 
\int_{D} d^2 x f^k A_k^a \epsilon^{ij}F^a_{ij} -\frac{\kappa}{4 \pi} 
\oint_{\partial D} d \varphi f^{\varphi} A^a_{\varphi} A^a_{\varphi},
\end{eqnarray}
and, imposing the $F^a_{ij}=0$, one is left with
\begin{eqnarray}
Q(f)=-\frac{\kappa}{4 \pi} \oint_{\partial D} d 
\varphi f^{\varphi} A^a_{\varphi} A^a_{\varphi},
\label{char}
\end{eqnarray}
where $A_{\varphi}=g^{-1}\partial_{\varphi} g$. Using the previous Poisson 
bracket (\ref{poisson}), we find
\begin{eqnarray}
\{ Q(f), Q(g) \} = Q([f,g]),
\end{eqnarray}
where $[f,g]\equiv g \partial_\varphi f -f \partial_\varphi g$.  
This is the Virasoro algebra {\it without} central term, called Witt 
algebra. So, if we restrict  the {\it Diff} along the circle ($\partial D$) 
there is no central term classically. The central term will arise only as 
a quantum mechanical effect of normal ordering. 


(b). From $A^a_r |_{\partial D}$ =constant, 
 $\partial_r A^a_{\varphi}|_{\partial D} =0$,
and ${A^a_0}|_{\partial D}\propto {A^a_\varphi}|_{\partial D}$,
(\ref{boundary}) 
becomes $\frac{dX}{dt}$ with 
$X=\frac{\kappa}{4 \pi} \oint_{\partial D} d \varphi 
f^r A^a_r A^a_{\varphi}$. 
The Noether charge becomes
\begin{eqnarray}
Q(f) &=&\frac{\partial L}{\partial \dot{A}^a_i} \delta_f A^a_i 
-X, \nonumber \\
&=&\frac{\kappa}{4 \pi} \int_{D} 
d^2 x f^k A^a_k \epsilon^{ij} F^a_{ij} -\frac{\kappa}{4 
\pi}
\oint _{\partial D} d \varphi (2 f^r A^a_r A^a_{\varphi} 
+f^{\varphi} A^a_{\varphi}A^a_{\varphi} ),
\end{eqnarray}
and, imposing $F^a_{ij}=0$, one is left with
\begin{eqnarray}
Q(f) = -\frac{\kappa}{4 \pi}
\oint _{\partial D} d \varphi 
(2 f^r A^a_r A^a_{\varphi} +f^{\varphi} A^a_{\varphi}A^a_{\varphi} ),
\label{cent}
\end{eqnarray}
where $A_{\varphi} =g^{-1}\partial_{\varphi} g$. Using the Poisson bracket 
(\ref{poisson}) and treating $A^a_r |_{\partial D}$ as a 
$c$-number, we find
\begin{equation}
\{ Q(f), Q(g) \} =Q([f,g]) +\frac{\kappa}{2 \pi} A^a_r A^a_r 
\oint_{\partial D} d \varphi f^r \partial_{\varphi} g^r. 
\label{virasoro}
\end{equation}
In general, this algebra does not satisfy the Jacobi identity 
and so the Noether charge $Q(f)$ as a symmetry generator can not be 
accepted. Therefore, the only way to avoid this undesirable situation is to 
consider the subset of transformation with particular 
$f^r|_{\partial D}\propto \partial _{\varphi} {f^{\varphi}|_{\partial D}}$
and $g^r|_{\partial D} 
\propto \partial _{\varphi} g^{\varphi}|_{\partial D}$ \cite{ref:Ban1}
such that only the  third order derivatives appear in the central term and 
hence (\ref{virasoro}) satisfies the Jacobi identity. 
A consatnt term $\propto~ \oint _{\partial D} d {\varphi} 
f^{\varphi} A^a_r A^a_r$ can be added to $Q(f)$ in (\ref{virasoro})
to obtain the first order 
derivative term in the center. 
Then, (\ref{virasoro}) 
becomes the standard form of the Virasoro algebra with central term by proper 
normalization of the proportionality constant \cite{ref:Ban1}.
So, in  contrast to 
the {\it Diff} along the circle ($\partial D$), {\it Diff}
 which 
deforms {\it across} the 
boundary has the central term even classically. 


Now, let us recompute the charges
after $F_{ij}^a=0$ was imposed from the beginning, and
compare with  previous results. 
Using $F_{ij}^a=0$, one can show that the second equation of  
(\ref{vari}) becomes \cite{ref:Jac3}
\begin{equation}
\delta_f A_i^a=D_i(f^iA_i)^a.
\end{equation}
Then by substituting this into
the variation of action (\ref{eq:2}) without the $F_{ij}^a$ term, we find
\begin{equation}
\delta_f L = \frac{d}{dt}  \left[\frac{\kappa}{2 \pi} \int_{D} d^2 x 
\epsilon^{ij} A_i^a D_j(f^kA_k)^a\right]+
\frac{\kappa}{2 \pi} \oint_{\partial{D}} d \varphi 
f^iA^a_i{\dot A}^a_\varphi.
\label{sympo}
\end{equation}
For {\it Diff} invariance, we demand the second 
term to be a total time derivative.
Again, we have two boundary conditions:
(a). $f^r|_{\partial D}=0$, and  (b). ${\dot{A}^a_r}|_{\partial D}=0$. 

(a). When $f^r|_{\partial D}=0$, we find the second term of
(\ref{sympo}) reduces to
\begin{equation}
\frac{d}{dt}  \left[\frac{\kappa}{4 \pi} \int_{D} d^2 x 
f^\varphi A^a_\varphi A^a_\varphi\right],
\end{equation}
and Noether charge is precisely the expression given in (\ref{char}).

(b). When $f^r|_{\partial D}\neq 0$, the second term can again be 
combined into a total 
derivative term by demanding that ${\dot{A}^a_r}|_{\partial D}=0$. 
Then,  we have the  charge precisely given in (\ref{cent}), 
and again one finds
the Virasoro algebra with central term. 
Note that unlike the derivation of (\ref{cent}), 
we need not to impose the extra 
conditions $\partial_r {A^a_\varphi}\vert_{\partial D}=0$,
and ${A^a_0}|_{\partial D}\propto {A^a_\varphi}|_{\partial D}$.

\section{Conclusion}

In summary, Kac-Moody and Virasoro algebra of  
Chern-Simons theory were derived 
by using the boundary symplectic structure 
which emerges as a consequence of symplectic reduction, 
and by applying the standard Noether procedures. 
The merit of our approach is that 
no assumptions about boundary charges are needed, 
and  the Kac-Moody algebra in its standard current density 
form (\ref{poisson}) is  obtained. 
It remains to be seen whether the Dirac's method
yields the same results. It would be also intersting
to extend the symplectic method  to the supersymmetric case
and to the higher dimensional 
Chern-Simons theories. 

Note added.-After completing this work, we became aware of 
Ref. \cite{ref:mick} in which the Kac-Moody algebra in Yang-Mills 
theory with Chern-Simons term was derived and generalized to 
higher dimensions. We thank J. Mickelsson for informing 
us about his work.
\section{Acknowledgements}

We would like to thank Prof. R. Jackiw for suggesting this problem
and guidance throughout the work. We also thank M. Ortiz for 
useful discussions. P.O. was supported by the Korea Ministry of 
Education through Basic Science Research Institute
(BSRI/98-1419) and by the Korea Science Engineering
Foundation (KOSEF) through the project number 
(95-0702-04-01-3). M.I.P. was supported by the financial support of 
Korea Research Foundation (KRF) made in the program year 1997.

\end{document}